\documentclass[11pt,a4paper]{article}
\pdfoutput=1
\usepackage{jcappub}
\usepackage{slashed}

\def\be{\begin{equation}}
\def\ee{\end{equation}}
\def\bea{\begin{eqnarray}}
\def\eea{\end{eqnarray}}

\notoc
\begin{document}

\title{Differential equation for partition functions and a duality pseudo-forest}

\author{Vitaly Vanchurin\footnote{E-mail: vvanchur@d.umn.edu}}

\date{\today}

\affiliation{Department of Physics, University of Minnesota, Duluth, Minnesota, 55812, USA \\
Duluth Institute for Advanced Study, Duluth, Minnesota, 55804, USA}

\abstract{We consider finite quantum systems defined by a mixed set of commutation and anti-commutation relations between components of the Hamiltonian operator. These relations are represented by an anti-commutativity graph which contains a necessary and sufficient information for computing the full quantum partition function. We derive a second-order differential equation for an extended partition function $z[\beta, \beta', J]$ which describes a transformation from a ``parent'' partition function $z[0, \beta', J]$  (or anti-commutativity graph) to a ``child'' partition function $z[\beta, 0, J]$  (or anti-commutativity graph). The procedure can be iterated and then one forms a pseudo-forest of duality transformations between quantum systems, i.e. a directed graph in which every vertex (or quantum system) has at most one incoming edge (from its parent system). The pseudo-forest has a single tree connected to a constant partition function, many pseudo-trees connected to self-dual systems and all other pseudo-trees connected to closed cycles of transformations between mutually dual systems. We also show how the differential equation for the extended partition function can be used to study disordered systems. }

\maketitle

\section{Introduction}\label{sec:introduction}

It is well known that the quantum partition function,  in principle, describes everything there is to know about the system, but, in reality, there are very few quantum systems for which  a closed from expression of the full partition function is known. For this reason it is often more productive to look for alternative representation of the partition functions other then the way it may be defined,
\be
{\cal Z}[\beta,J] = tr\left [\exp \left (- \beta \hat{H}  + \sum_{i=1}^K  J_i \hat{\cal O}^i \right )\right ].\label{eq:partition}
\ee
One (gigantic) framework for modeling partition functions is known as the effective field theories \cite{Zee} which boils down to representing ${\cal Z}[\beta,J]$ as a path integral over some effective fields 
\be
{\cal Z}[\beta,J] = \int {\cal D} \varphi   \;\exp\left( {-\beta {\cal S}[\varphi] + \sum_{i=1}^K  J_i {\cal O}^i[\varphi] }\right )\label{eq:effective}
\ee
where by $\varphi$  we abbreviate all fields at all points. The representation proved very useful not only for developing the standard model of particle physics, but also for many non-perturbative calculations in context of quantum field theories \cite{Shifman}. Moreover, the path integral representations with emergent additional dimensions were indispensable for finite temperature calculations in context of the imaginary time formalism \cite{Kamenev } and for quantum gravity calculations in context of the AdS/CFT correspondence \cite{Maldacena, Witten}. There is certainly a lot more to be done in that direction, but in this article we will take a somewhat different approach which was largely motivated by our results in Ref. \cite{Vanchurin}. 

Instead of using path integrals, we will try to identify a differential equation which is solved by the quantum partition function \eqref{eq:partition}. In a sense we will be developing an {\it operator representation} of the partition functions by identifying differential operators (not to confuse with quantum operators) which annihilate ${\cal Z}[\beta,J]$ or by deriving differential equations (and the corresponding boundary/initial conditions) which are solved by ${\cal Z}[\beta,J]$, 
\be
\hat{L} \;z[\beta, \beta', J]  = 0\;\;\;\;\text{where}\;\;\;\; z[\beta, 0, J] = {\cal Z}[\beta,J].\label{eq:operator}
\ee
For the systems considered in this paper the partition function will be extended into extra dimension $\beta'$, the differential operator $\hat{L}$ will be of second order in $\beta$ (but first order in $\beta'$) and the appropriate initial conditions at $\beta=0$ will be set by another partition function, which we shall call the ``parent'' partition function (see Sec. \ref{sec:extended}).  Note, that in the functional renormalization group (RG) computations one also solves a differential equation for Schwinger functional $W[J] = \log({\cal Z}[\beta,J])$ \cite{Polchinski} or for effective action $\Gamma[\varphi]$ (i.e. Legendre transform of $\log({\cal Z}[\beta,J])$)  \cite{Wetterich, Morris}, but the equation is first-order in RG scale and the initial conditions are set by the action at the microscopic scales. In our case the differential equation will necessarily be of second order and the partition function will be extended into extra dimension described by $\beta'$. Nevertheless, it would be interesting to see if there is a connection to the RG calculations that we have not yet identified. 

Another important byproduct of our analysis is that the differential equation for the extended partition functions \eqref{eq:operator} can be viewed as a duality transformations from parent systems to children systems. If one represents different quantum systems as vertices of a graph and the duality transformations as directed edges of the graph, then the resulting graph can be shown to be a directed pseudo-forest, i.e. a graph in which every vertex, or a system, has at most one incoming edge from its parent system (see Sec. \ref{sec:forest}). The pseudo-forest does not include all of the quantum systems, but only those for which the quantum partition function is symmetric under a sign flip, i.e. ${\cal Z}[\beta,J] = {\cal Z}[-\beta,J]$. Such systems are somewhat simpler as they can be defined by a mixed set of commutation and anti-commutation relations between $\hat{H}$ and $\hat{{\cal O}}^i$'s operators (see Sec. \ref{sec:tracelessness}). We shall also discuss how the formalism can be used to study disordered systems where individual realizations of the partition function may not be symmetric, but the disorder-averaged partition function is such that all odd statistical moments vanish (see Sec. \ref{sec:disordered}). 

The paper is organized as follows.  In the next section we define a set of gamma operators which will be the building blocks for constructing different quantum systems. In Sec. \ref{sec:anti-commutativity} we define the anti-commutativity graph which carries information about commutation and anti-commutation relations between gamma operators. In Sec. \ref{sec:tracelessness} we impose a tracelessness condition on the gamma operators which restricts the analysis to only symmetric partition functions. In Sec. \ref{sec:extended} we derive a differential equations for the extended partition function which maps the partition function of a parent system to the partition function of a child system. The procedure is illustrated in Sec. \ref{sec:examples} with three different examples and then discussed in Sec. \ref{sec:disordered} in context of disordered systems. In Sec. \ref{sec:forest} we show how the duality map can be iterated to define a duality pseudo-forest. In Sec.  \ref{sec:discussion} we summarize and discuss the main results of the paper. 

\section{Gamma operators}\label{sec:operators}

Consider a quantum partition function 
\be
{\cal Z}[\beta, J] = tr\left [\exp \left (\beta \sum_{i=1}^K  J_i \hat{\Gamma}^i \right )\right ],\label{eq:partition2}
\ee
where all of the gamma operators are Hermitian
\be
\hat{\Gamma}^i =\hat{\Gamma}^{i \dagger}  \label{eq:Hermitian}
\ee
 and the coefficients $J_i$'s are real numbers. This partition function is equivalent to \eqref{eq:partition} after appropriate redefinition $K+1 \rightarrow K$ and substitution, e.g. 
 \bea
 \hat{\Gamma}^i &=& \frac{1}{\beta} \hat{\cal O}^i,\\ 
 \hat{\Gamma}^{K} &=&- \frac{1}{J_{K}} \hat{H}.
 \eea
 From now on, we shall refer to the sum of all terms in the exponent in \eqref{eq:partition2} as the Hamiltonian operator,
\be
\hat{\cal H}[J] = \sum_{i=1}^K J_i \hat{\Gamma}^i \label{eq:Hamiltonian}
\ee
whether or not some of the coefficients $J_i$'s  would be set to zero at the end of the calculations as is often the case for sources. The main advantage of viewing the Hamiltonian operator as a function of many coefficients \eqref{eq:Hamiltonian} is that it allows us to represent/parametrize many different quantum theories by a single function of coefficients $J_i$'s. 

For a given set of $\hat{\Gamma}^i$ operators, the corresponding theory subspace (spanned by real coordinates $J_i$'s) can be shown to be an inner product space with respect to operator addition, i.e.
\be
\sum_i  A_i \hat{\Gamma}^i +  \sum_i B_i \hat{\Gamma}^i = \sum_i (A_i + B_i) \hat{\Gamma}^i
\ee
and scalar product defined by the trace, i.e.
\be
\left (\sum_i  A^i \hat{\Gamma}_i,  \sum_j B_j \hat{\Gamma}^j  \right ) \equiv {\cal N}^{-1} \sum_{i,j}  tr[ (A^i \hat{\Gamma}_i)  (B_j \hat{\Gamma}^j)] 
\ee
where 
\bea
{\cal N} &=& tr\left [ \hat{I} \right ],\\
A_i&=&A^i,\\
\hat{\Gamma}^i &=& \hat{\Gamma}_i.
\eea
Then, we can rewire the Hamiltonian operator in orthonormal basis, but to avoid unnecessary complications we shall assume that the original $\hat{\Gamma}^i$ operators are already orthonormal
\be
\left (\hat{\Gamma}_i, \hat{\Gamma}^j \right ) =  \delta^{ij}\label{eq:orthonormality}
\ee 
and then
\be
\left (\sum_i  A^i \hat{\Gamma}_i,  \sum_j B_j \hat{\Gamma}^j \right ) =\sum_{i,j} A^i B_j \; {\cal N}^{-1}  tr[\hat{\Gamma}_i \hat{\Gamma}^j]  = \sum_i A^i B_i.
\ee

There is still a remaining freedom of choosing the orthonormal basis which can be used to set all but one operator, i.e. the identity operator $\hat{\Gamma}^0 = \hat{I}$,  to be traceless 
\be
 \delta^{i0} = \left (\hat{\Gamma}^i, \hat{\Gamma}^0 \right ) = {\cal N}^{-1}  tr[ \hat{\Gamma}^i  \hat{\Gamma}^0] =  {\cal N}^{-1}  tr[ \hat{\Gamma}^i]
\ee
and to square to identity
\be
\left ( \hat{\Gamma}^i\right )^2 = \hat{I}. \label{eq:square}
\ee
The procedure can be made more transparent by expressing gamma operators in terms of tensor products of Pauli operators $\hat{\sigma}^1, \hat{\sigma}^2, \hat{\sigma}^3$  and identity $\sigma^0$, i.e.
\be
\hat{\Gamma}^i = \hat{\sigma}^{i_1} \otimes  \hat{\sigma}^{i_2}  \otimes ...  \otimes \hat{\sigma}^{i_N} =  \bigotimes_{n=1}^N  \hat{\sigma}^{i_n}  \label{eq:Pauli}
\ee
where $i$ is an integer in base four with $N$  digits denoted by $i_n \in \{ 0,1,2,3\}$. Such operators are guaranteed to be traceless (with an exception of identity $ \hat{\Gamma}^0 = \hat{I}$), and the inclusion of the identity produces a partition function which is related to the one without identity by a trivial transformation 
\be
{\cal Z}[\beta, J] = tr\left [\exp \left (\beta \sum_{i=0}^K  J_i \hat{\Gamma}^i \right )\right ] = e^{\beta {J_0} }  tr\left [\exp \left (\beta \sum_{i=1}^K  J_i \hat{\Gamma}^i \right )\right ]. 
\ee 
For this reason and without loss of generality we shall assume that $J_0=0$ (unless states otherwise) and then all of gamma operators in the Hamiltonian \eqref{eq:Hamiltonian} are traceless,
\be
 tr[ \hat{\Gamma}^i] = 0 \label{eq:traceless}. 
\ee

\section{Anti-commutativity graph}\label{sec:anti-commutativity}

If the gamma operators either commute or anti-commute, then we can define an adjacency matrix
\bea
{ {\cal A}^{ij} }= \begin{cases} 0 \;\; \;\;\;\;\text{if} \;\;\;\;\;\; [\hat{\Gamma}^{i}, \hat{\Gamma}^{j}] \equiv \hat{\Gamma}^{i}  \hat{\Gamma}^{j} - \hat{\Gamma}^{j} \hat{\Gamma}^{i} = 0 \\
1 \;\;\;\;\;\;  \text{if} \;\;\;\;\;\;  \{\hat{\Gamma}^{i}, \hat{\Gamma}^{j}\} \equiv \hat{\Gamma}^{i}  \hat{\Gamma}^{j} + \hat{\Gamma}^{j}\hat{\Gamma}^{i}  =0  \end{cases} \label{eq:adjacency}
\eea
of a graph with nodes representing individual operators, i.e. $\hat{\Gamma}^i$'s,  and edges representing pairs of anti-commuting operators,  i.e. $\{\hat{\Gamma}^{i}, \hat{\Gamma}^{j}\} = 0$. We will refer to the graph as an anti-commutativity graph, but one could have also defined a complement graph (i.e. commutativity graph) without loosing generality. Note that even if the original gamma operators do not satisfy the condition \eqref{eq:adjacency},  it is always possible to decompose them into linear sums of orthonormal operators (such as generalized Pauli operators \eqref{eq:Pauli}) for which the condition is satisfied.  

\begin{figure}[]
\begin{center}
\includegraphics[width=0.8\textwidth]{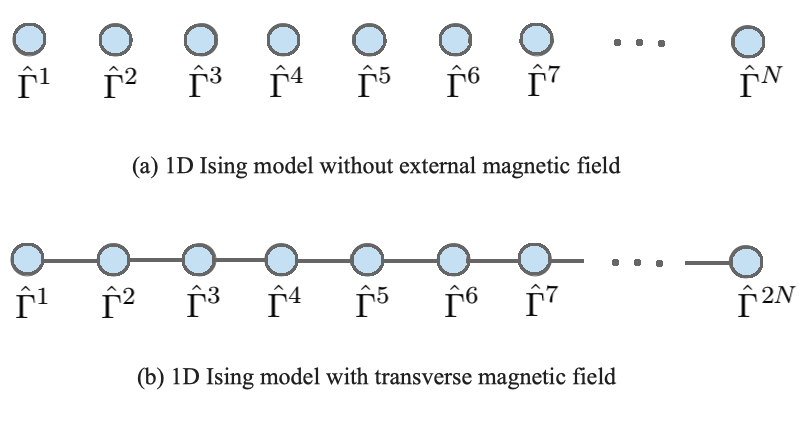}
\caption{Anti-commutativity graphs for $1D$ Ising models} \label{fig:ising}
\end{center}
\end{figure}

Consider a one dimensional Ising model with Hamiltonian given by
\be
\hat{\cal H} [J]= \sum_{i=1}^N  J_i  \hat{\Gamma}^i =  \sum_{n=1}^N J_n \hat{\sigma}_n^3\hat{\sigma}_{n+1}^3
\ee
where
\be
\hat{\sigma}^a_n \equiv  \left ( \bigotimes_{m=1}^{n-1}  \hat{\sigma}^0 \right ) \otimes \hat{\sigma}^a \otimes \left ( \bigotimes_{m=n+1}^{N}  \hat{\sigma}^0 \right ) 
\ee
and
\be
\hat{\Gamma}^i = \hat{\sigma}_i^3\hat{\sigma}_{i+1}^3.
\ee
Since all of the gamma operators commute, the adjacency matrix is ${ {\cal A}^{ij} }=0$ and the anti-commutativity graph contains $N$ nodes with no edges  (see Fig. \ref{fig:ising}.a) 
However, if we add a transverse magnetic field then the Hamiltonian is given by
\be
\hat{\cal H} [J]=  \sum_{i=1}^{2N}  J_i  \hat{\Gamma}^i =   \sum_{n=1}^{N} J_{2n} \hat{\sigma}_{n}^3\hat{\sigma}_{n+1}^3 + \sum_{n=1}^{N} J_{2n-1} \hat{\sigma}_{n}^1
\ee
where 
\bea
 \hat{\Gamma}^{2i} &=&  \hat{\sigma}_{i}^3\hat{\sigma}_{i+1}^3 \notag\\
  \hat{\Gamma}^{2i-1} &=&  \hat{\sigma}_{i}^1, \label{eq:Ising_gamma}
\eea
and the corresponding anti-commutativity graph is shown on Fig. \ref{fig:ising}.b. Note that in the limit of infinite lattice (to both positive and negative infinities) the anti-commutativity graph is symmetric under a ``shift'' of vertices $i\rightarrow i+1$. This symmetry is responsible for the well-known self-duality between low and high temperatures Ising models, but clearly the anti-commutativity graph has many more symmetries. For example, the graph is invariant under ``reflection'' about $k$th vertex, i.e. $i \rightarrow 2k - i$, or about $(k,k+1)$ edge, $i \rightarrow 2k+1 - i$.

\begin{figure}[]
\begin{center}
\includegraphics[width=1\textwidth]{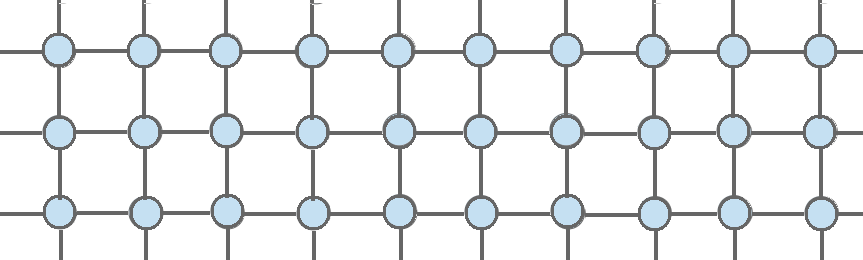}
\caption{Anti-commutativity graph for $2D$ Ising model with transverse magnetic field.} \label{fig:ising2}
\end{center}
\end{figure}

The situation is not very different in higher dimensions. For example, in two-dimensions the anti-commutativity graph of the Ising model with transverse magnetic field is shown on Fig. \ref{fig:ising2}. Evidently, there is still a shift symmetry of the graph which is responsible for the well-known Kramers-Wannier duality \cite{KW}, but many other transformations would leave the anti-commutativity graph invariant. In fact, the group of symmetry transformations is now much larger since in addition to ``shifts'' and  ``reflections'', the graph is also invariant under discrete ``rotations'' (i.e.  by $\pi/2$, $\pi$ and $3\pi/2$) around any of the vertices. 

In general, the symmetries of an anti-commutativity graph can be expressed as 
\be
{\cal A} = P^{\dagger} {\cal A} P \label{eq:self-duality} 
\ee
where $P$ is some permutation matrix which represents permutations of vertices of the graph and  $P^\dagger$ is its transpose. For example, the anti-commutativity for $1D$ Ising model with transverse magnetic field and periodic boundary conditions is given by
\be
{\cal A}^{ij} = \begin{cases} 1 \;\;\;&\text{if} \;\;\;  i=j+1 \mod N \\
1 \;\;\;&\text{if} \;\;\;  j=i+1 \mod N \\
0 \;\;\;&\text{otherwise}
 \end{cases} 
\ee
and the self-duality condition \eqref{eq:self-duality} is respected for a shift matrix
\be
P_{ij} = \begin{cases} 1 \;\;\;&\text{if} \;\;\;  i=j+1 \mod N \\
0 \;\;\;&\text{otherwise}
 \end{cases} 
\ee
Then the self-duality of the Ising model can be expressed as \eqref{eq:self-duality}, but now the duality is not exact due to global (or topological) effects. At this point one might wonder if there exists an additional condition that the gamma operators could obey in order for the symmetries of the anti-commutativity graph to translate into exact symmetries of the partition functions or to self-dualities of the corresponding systems. In the following section we will discuss one such condition which will allow  us to identify the self-dualities as well as more general duality transformations. 

\section{Tracelessness condition}\label{sec:tracelessness}

So far we assumed that the gamma operators are Hermitian \eqref{eq:Hermitian}, orthonormal \eqref{eq:orthonormality},  square to identity \eqref{eq:square}, traceless \eqref{eq:traceless} and either commute or anti-commute \eqref{eq:adjacency}. These assumption are not restrictive in a sense that one can always decompose any Hamiltonian  \eqref{eq:Hamiltonian} into a linear sum of $\hat{\Gamma}^i$'s which satisfy these condition. In this section we will make an actually restrictive assumption which will not allow us to study all possible quantum systems, but will allow us to extract self-dualities from the symmetries of the anti-commutativity graph and to derive an equation which the corresponding quantum partition functions must satisfy. The assumption can be viewed as a generalization of both normality \eqref{eq:orthonormality} and tracelessness \eqref{eq:traceless} conditions to arbitrary products of gamma operators,
\be
tr\left [\hat\Gamma^{i_1} \hat\Gamma^{i_2} ... \hat\Gamma^{i_k} \right ] = 0 \label{eq:tracelessness} 
\ee
where $i_1 <i_2 < ... <i_k$. From \eqref{eq:traceless}  we already know that a single  $\hat{\Gamma}^i$ is traceless, and from \eqref{eq:orthonormality}  that a product of two different $\hat{\Gamma}^i$'s is traceless, but the condition \eqref{eq:tracelessness} implies that an arbitrary product of different $\hat{\Gamma}^i$'s is traceless which  is certainly more restrictive. 

If we expand the partition function \eqref{eq:partition2} in a power series, 
\bea
{\cal Z}[\beta, J] &=& \sum_{n=0}^\infty \frac{\beta^n}{n!} tr\left [\left (\sum_{i=1}^K  J_i \hat{\Gamma}^i \right )^n\right ] \notag\\
&=& \sum_{n=0}^\infty \frac{\beta^n}{n!}  \sum_{i_1=1}^K  \sum_{i_2=1}^K  ...  \sum_{i_n=1}^K  J_{i_1}J_{i_2} ... J_{i_n} tr\left [ \hat{\Gamma}^{i_1} \hat{\Gamma}^{i_2} ... \hat{\Gamma}^{i_n} \right ]    \label{eq:partition3}
\eea
then the tracelessness condition \eqref{eq:tracelessness} implies that the traces of all terms with at least one odd power of $\hat{\Gamma}^{i}$'s (or equivalently of $J_{i}$'s) would vanish. On the other hand all of the terms with only even powers of $J_{i}$'s would be multiplied by a trace of either plus or minus identity, i.e.
\be
tr\left [ \hat{\Gamma}^{i_1} \hat{\Gamma}^{i_2} ... \hat{\Gamma}^{i_n} \right ]  = \begin{cases} +tr\left[ \hat{I} \right ] = + {\cal N} \;\;\;&\text{gamma operators can be ordered with an}\vspace*{-0.1in} \\ 
& \text{even number of anti-commutation relations,} \\  
-tr\left[ \hat{I} \right ] = - {\cal N} \;\;\;&\text{gamma operators can be ordered with an}\vspace*{-0.1in} \\&
\text{odd number of anti-commutation relations.} \end{cases} \label{eq:even_moments}
\ee
But since the sign is given by the number of times the anti-commutativity relations are used to order the gamma operators, it is uniquely determined by the adjacency matrix ${\cal A}^{ij}$. This means that if the tracelessness condition is satisfied, then the partition function is in one-to-one correspondence with the anti-commutativity graph: for every anti-commutativity graph there is a partition function and for every partition function there is an anti-commutativity graph.  Therefore, if there is a symmetry transformation which leaves an anti-commutativity graph invariant \eqref{eq:self-duality}, then there must be a self-duality transformation which leaves the corresponding partition function invariant and vise versa. For example, the gamma operators in the Ising models with transverse magnetic field \eqref{eq:Ising_gamma} do satisfy the tracelessness conditions \eqref{eq:tracelessness}  and so it should not be surprising that  the ``shift'' symmetry of the anti-commutativity graph corresponds to the  Kramers-Wannier duality \cite{KW}. There are however other symmetries of the anti-commutativity graph (i.e. ``reflection'' and``rotation'') which can be viewed as more general self-dualities of the Ising model. 

In summary, the gamma operators are assumed to be (a) Hermitian \eqref{eq:Hermitian}, (b) square to identity \eqref{eq:square} and to satisfy the (c) commutativity \eqref{eq:adjacency} and (d) tracelessness \eqref{eq:tracelessness} conditions, and the only restrictive condition (i.e. tracelessness condition) implies that all odd statistical moments vanish or that the partition function is symmetric under a sign flip, i.e. 
\bea
 {\cal  Z}[\beta, J] &=&  {\cal  Z}[-\beta, J] \label{eq:symmetric}\\
{\cal  Z}[\beta, J_1, ... ,J_i, ... J_K] &=& {\cal  Z}[\beta, J_1, ... ,-J_i, ... ,J_K] \;\;\;\;\;\;\;\;\forall i \in \{1,...,K\}. \notag
\eea

\section{Extended partition function}\label{sec:extended}

If the gamma operators satisfy the tracelessness condition  \eqref{eq:tracelessness} then equation \eqref{eq:symmetric} can be used to rewrite the partition function as
\bea
{\cal  Z}[\beta, J] &=&  tr\left[\exp\left(\beta \hat{\cal H}\right ) \right ] \notag \\
&=& tr\left[\cosh\left(\beta  \hat{\cal H} \right ) \right ] \notag \\
&=& tr\left[\cosh\left(\beta  \sqrt{\hat{\cal H}^2} \right ) \right ]\notag \\
&=& tr\left[\cosh\left(\beta  \sqrt{ \sum_{i,j} J_i J_j \hat\Gamma^i \hat\Gamma^j } \right ) \right ].\label{eq:old_partition}
\eea
Then there are only  four possibilities for the individual terms inside of the square-root, 
\be
\hat\Gamma^i \hat\Gamma^j = \begin{cases} \hat{I} \;\;\;\; &\text{if} \;\;\; i=j \\
0 \;\;\;\; &\text{if} \;\;\;  i \neq  j \;\;\text{and}\;\;\; {\cal A}^{ij} = 1 \\
\hat\Gamma^{ij}  \;\;\;\; &\text{if} \;\;\;  i < j  \;\;\text{and}\;\;\; {\cal A}^{ij} = 0 \\
\hat\Gamma^{ji}  \;\;\;\; &\text{if} \;\;\;  i > j \;\;\text{and}\;\;\; {\cal A}^{ij} = 0 
\end{cases}\label{eq:new_operators}
\ee
where in the third and forth cases we simply defined the combined operators $\hat\Gamma^{ij}$'s from pairs of commuting operators with the first index (by convention) smaller than the second one. If we introduce an auxiliary partition function for the combined operators, i.e.
\be
{\cal Z}'[\beta', P] = tr\left [ \exp\left (\beta' \sum_{i<j}  P_{ij} \hat\Gamma^{ij} \right )\right ],
\ee
then the equation \eqref{eq:old_partition} can be used to express the original partition function ${\cal Z}[\beta, J_i]$ in terms of ${\cal Z}'[\beta', J_i J_j]$, i.e.
\bea
{\cal Z}[\beta, J_i] &=&  tr\left[\cosh\left(\beta  \sqrt{ \sum_{i,j} J_i J_j \hat\Gamma^i \hat\Gamma^j } \right ) \right ] \notag\\
&=&  tr\left[\cosh\left(\beta  \sqrt{\sum_i J_i^2 \hat{I} +  2 \sum_{i<j} J_i J_j \hat\Gamma^{ij} } \right ) \right ]\notag\\
 &=& \cosh\left(\beta  \sqrt{ \sum_i J_i^2  +  2 \frac{\partial}{\partial \beta'} } \right ) {\cal Z}'[\beta', J_i J_j]_{\beta'=0}.\label{eq:equation}
\eea
This equation already describes a map from a ``parent'' system (or partition function ${\cal Z}'[\beta', J_i J_j]$ to a ``child'' system (or partition function ${\cal Z}[\beta, J]$), but the map is non-local in a sense that arbitrary high derivatives of the parent partition function are needed to determine the child partition function. 

The problem can be solved by defining an extended partition function
\be
z[\beta, \beta', J] \equiv \cosh\left(\beta   \sqrt{ \sum_i J_i^2 + 2 \frac{\partial}{\partial \beta' }} \right ) {\cal Z}'[\beta', J_i J_j]
\ee
which combines together both the parent and the child partition functions. Then, if the parent function ${\cal Z}'[\beta', J_i J_j] = z[0,\beta',  J]$ is known, the child function ${\cal Z}[\beta, J]  = z[\beta,0,  J] $ can be obtained by solving the following differential equation 
\be
\boxed{ \left (\frac{\partial^2}{ \partial \beta^2} - \sum_i J_i^2  - 2 \frac{\partial}{\partial \beta'}  \right ) z[\beta, \beta', J] = 0 }\label{eq:the-equation}
\ee
with initial conditions
\bea
 z[0,\beta',  J]  &=& {\cal Z}'[\beta', J_i J_j]\\
\frac{\partial  }{ \partial \beta}  z[0,\beta',  J]  &=& 0.
\eea
(See Fig \ref{fig:extended} for plots of a sample extended partition function \eqref{eq:mixed-solution}.)

Likewise, if ${\cal Z}[\beta, J]  = z[\beta,0,  J] $ is known, then ${\cal Z}'[\beta', J_i J_j] = z[0,\beta',  J]$ can be obtained by solving the same equation \eqref{eq:the-equation} but in ``orthogonal'' direction with initial conditions
\be
z[\beta, 0, J] =  {\cal Z}[\beta, J].
\ee
Note, however, that the inverse map (from a child system to a parent system) does not tell us everything about the parent partition function ${\cal Z}'[\beta', P]$, but only on how it depends on factorizable sources $P_{ij} = J_i J_j$. Nevertheless, this will be sufficient for the analysis of disordered systems in Sec. \ref{sec:disordered}. 

\section{Examples}\label{sec:examples}

\begin{figure}[]
\begin{center}
\includegraphics[width=1\textwidth]{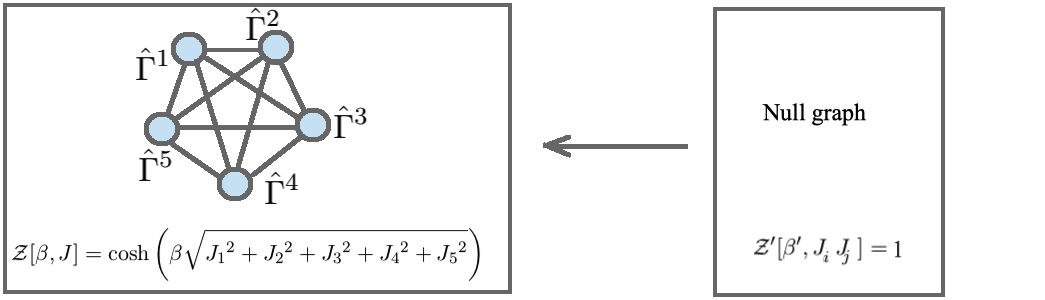}
\caption{Duality map from a constant parent partition function to a child partition function described by a complete anti-commutativity graph with $K=5$ gamma operators.} \label{fig:complete}
\end{center}
\end{figure}

To illustrate the procedure, that was introduced in the previous section, consider a quantum Hamiltonian operator \eqref{eq:Hamiltonian} with all of the gamma operators anti-commuting, i.e.
\be
\{ \hat{\Gamma}^i, \hat{\Gamma}^j \} = \delta^{ij}.
\ee
The corresponding anti-commutativity graph is a complete graph and the partition function can be obtained by solving equation \eqref{eq:the-equation} with initial conditions
\bea
 z[0,\beta',  J]  &=& 1 \\
\frac{\partial  }{ \partial \beta}  z[0,\beta',  J]  &=& 0.\notag
\eea
The solutions is given by
\be
z[\beta,\beta', J] = \cosh\left(\beta \sqrt{\sum_{i} {J_i}^2}\right) 
\ee
and thus the partition function is
\be
{\cal Z}[\beta, J]  = z[\beta,0, J] = \cosh\left(\beta \sqrt{\sum_{i} {J_i}^2}\right). 
\ee
See Fig. \ref{fig:complete}  for an illustration of the duality map from a constant partition function (i.e. null anti-commutativity graph) to a system with $K=5$ anti-commuting gamma operators (i.e. complete anti-commutativity graph).

\begin{figure}[]
\begin{center}
\includegraphics[width=1.15\textwidth]{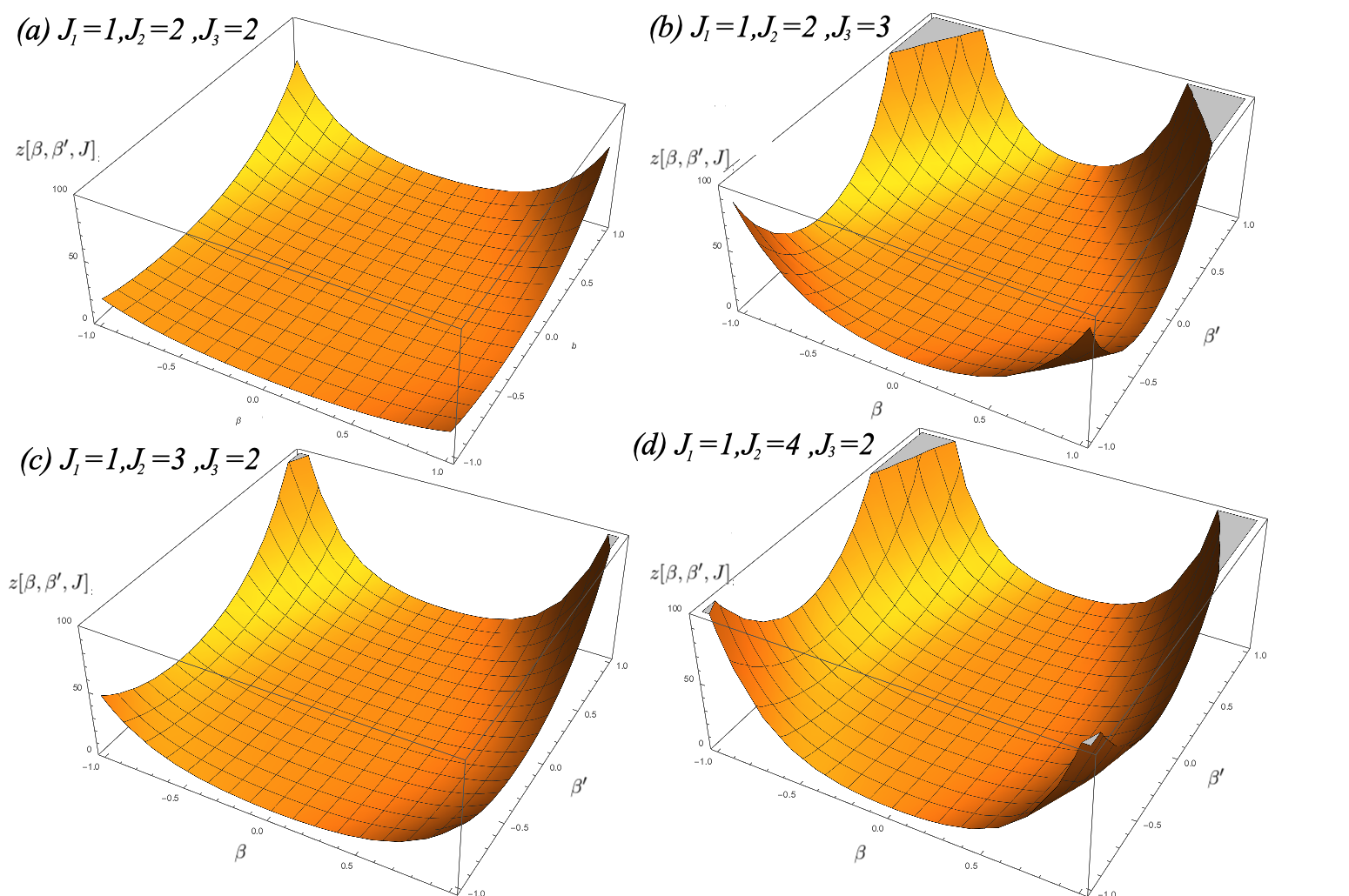}
\caption{Extended partition function $z[\beta, \beta',  J_1, J_2, J_3]$ for four different sets of values of $J_i$'s coefficients.  } \label{fig:extended}
\end{center}
\end{figure}

\begin{figure}[]
\begin{center}
\includegraphics[width=1.05\textwidth]{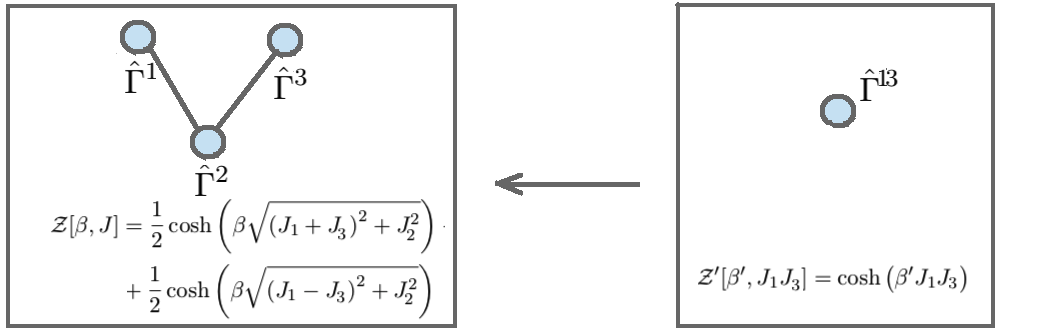}
\caption{Duality map from a parent partition function with a single gamma operator to a child partition function described by three gamma operators satisfying  commutation and anti-commutation relations \eqref{eq:mixed}.} \label{fig:mixed}
\end{center}
\end{figure}

The second example is a system described by a Hamiltonian \eqref{eq:Hamiltonian} with three gamma operators which satisfy the  following commutation and anti-commutation relations, 
\bea
\{ \hat{\Gamma}^1, \hat{\Gamma}^2 \} &=& 0\notag\\
\{ \hat{\Gamma}^2, \hat{\Gamma}^3 \} &=& 0\label{eq:mixed}\\
\left [ \hat{\Gamma}^1, \hat{\Gamma}^3 \right ] &=& 0.\notag 
\eea
Now, there is one combined operator, i.e. $\hat{\Gamma}^{13}=\hat{\Gamma}^1 \hat{\Gamma}^3$, whose partition function sets the initial conditions 
\bea
z[0,\beta',J] &=& {\cal Z}'[\beta',J_1 J_3] = \cosh\left(\beta' J_1 J_3\right) \label{eq:mixed-initial}  \\
\frac{\partial  }{ \partial \beta}  z[0,\beta',  J]  &=& 0.\notag
\eea
for  the differential equation \eqref{eq:the-equation}, i.e.
\be
 \left (\frac{\partial^2}{ \partial \beta^2} - \left (J_1^2 + J_2^2 + J_3^2\right )  - 2 \frac{\partial}{\partial \beta'}  \right ) z[\beta, \beta', J] = 0. \label{eq:mixed-equation}
\ee
Then the solution for the extended partition function is given by
\be
z[\beta,\beta', J] =\frac{1}{2}  e^{\beta' J_1 J_3} \cosh\left(\beta \sqrt{ \left ( J_1 + J_3\right )^2 + J_2^2}\right) + \frac{1}{2} e^{- \beta' J_1 J_3} \cosh\left(\beta \sqrt{ \left ( J_1 - J_3\right )^2 + J_2^2}\right) \label{eq:mixed-solution}
\ee
(plotted on Fig. \ref{fig:extended}) and the child partition function is
\be
{\cal Z}[\beta, J]  = z[\beta,0, J] = \frac{1}{2} \cosh\left(\beta \sqrt{ \left ( J_1 + J_3\right )^2 + J_2^2}\right) + \frac{1}{2}  \cosh\left(\beta \sqrt{ \left ( J_1 - J_3\right )^2 + J_2^2}\right).
\ee
The corresponding duality map is illustrated on Fig. \ref{fig:mixed}. 

\begin{figure}[]
\begin{center}
\includegraphics[width=1\textwidth]{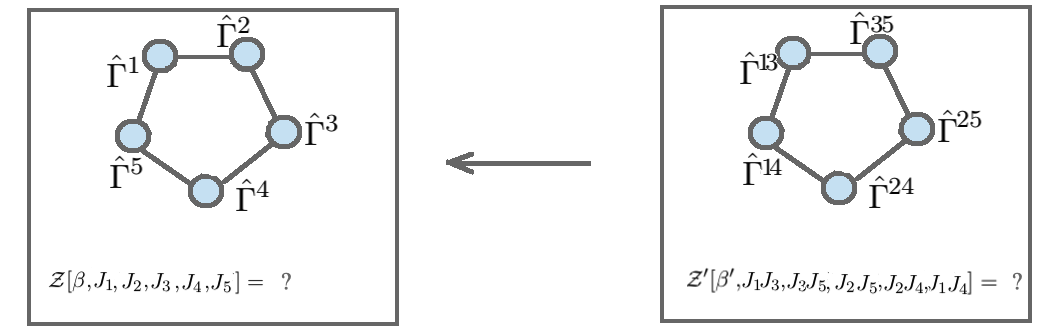}
\caption{Self-duality map in a system with $K=5$ gamma operators satisfying the commutation relations described by a pentagon anti-commutativity graph.} \label{fig:selfdual}
\end{center}
\end{figure}

The third example is a self-duality map where the gamma operators in both parent and child partition functions satisfy the same commutation and anti-commutation relations described by the anti-commutativity graphs that are shown on Fig. \ref{fig:selfdual}. We can still set-up an equation for the extended partition function, i.e.
\be
 \left (\frac{\partial^2}{ \partial \beta^2} - \left (J_1^2 + J_2^2 + J_3^2+ J_4^2+ J_5^2\right )  - 2 \frac{\partial}{\partial \beta'}  \right ) z[\beta, \beta', J] = 0, \label{eq:-dualequation}
\ee
but now the initial conditions depend on the solution itself (i.e. self-duality), 
\bea
z[0,\beta',J_1, J_2, J_3, J_4, J_5] &=& z[\beta',0 ,J_1 J_3, J_3 J_5, J_2 J_5, J_2 J_4, J_1 J_4]  \label{eq:self-dual-initial}  \\
\frac{\partial  }{ \partial \beta}  z[0,\beta',  J]  &=& 0.\notag
\eea
This is certainly a complication in comparison with previous examples, but one can try to solve it using iterative methods, e.g. stating with a constant initial condition $z_1[0,\beta',J]=1$ to solve for $z_1[\beta,\beta',J]$ and then use it to improve the initial conditions $z_2[0,\beta',J]= z_1[\beta',0,J]$, etc. In the limit of a large number of steps the iterative procedure is expected to converge to the true partition function, i.e. ${\cal Z}[\beta, J] = \lim_{n\rightarrow \infty} z_n[\beta,0,J]$.

\section{Disordered systems}\label{sec:disordered}

Before we switch to the discussion of dualities, let us briefly show how the formalism can be used to study disordered systems. As was already mentioned the only restrictive condition that we imposed on the gamma operators was that they must satisfy the tracelessness condition \eqref{eq:tracelessness}. What this means is that in the power series expansion of the partition function \eqref{eq:partition3} the traces of all terms with at least one odd power of gamma operator must vanish. This is a generic situation in disordered systems where some of the parameters  of the model (such as $J_{i}$'s) are viewed as random variables. For example, if all of $J_{i}$ coefficients are drawn from a symmetric probability distribution, such as Gaussian, then all odd statistical moments must vanish. Then even if the tracelessness condition \eqref{eq:tracelessness} is not satisfied by operators, the respective terms in the disorder-averaged partition function $\langle {\cal  Z}[\beta, J] \rangle$ would still vanish if all odd statistical moments vanish, i.e.
\be
\langle  J^{2n +1}_{i} \rangle = 0,\label{eq:odd_moments}
\ee
for all integers $i$ and $n$. To illustrate this, consider expansion of the partition function \eqref{eq:partition3} averaged over disorder,
\bea
 {\cal Z}[\beta, J] &=& \sum_{n=0}^\infty \left  \langle \frac{\beta^n}{n!} tr\left [\left (\sum_{i=1}^K  J_i \hat{\Gamma}^i \right )^n\right ] \right \rangle \notag\\
&=& \sum_{n=0}^\infty \frac{\beta^n}{n!}  \sum_{i_1=1}^K  \sum_{i_2=1}^K  ...  \sum_{i_n=1}^K  \left \langle J_{i_1}J_{i_2} ... J_{i_n} \right \rangle tr\left [ \hat{\Gamma}^{i_1} \hat{\Gamma}^{i_2} ... \hat{\Gamma}^{i_n} \right ]  \label{eq:part_averaged}
\eea
As in Sec. \ref{sec:extended} the odd moments would still vanish, but not because of the tracelessness condition, but because of the disorder described by \eqref{eq:odd_moments}. This implies that the disorder-averaged partition function must still be symmetric under a sign flip \eqref{eq:symmetric} and thus can be expressed as
\bea
{\cal  Z}[\beta, J]  &=& \left  \langle tr\left[\exp\left(\beta \hat{\cal H}\right ) \right ]   \right \rangle = \left  \langle  tr\left[\cosh\left(\beta  \hat{\cal H} \right ) \right ]  \right \rangle = \left  \langle   tr\left[\cosh\left(\beta  \sqrt{\hat{\cal H}^2} \right ) \right ]  \right \rangle \notag \\
&=&\left  \langle   tr\left[\cosh\left(\beta  \sqrt{ \sum_{i,j} J_i J_j \hat\Gamma^i \hat\Gamma^j } \right ) \right ] \right \rangle \label{eq:avg_partition}
\eea
which is equivalent to \eqref{eq:old_partition}. As a result, the disorder-averaged partition function can still be obtained by solving the differential equation of an extended partition function \eqref{eq:the-equation}. 

For example, consider an exactly solvable disordered system of Majorana fermions with interactions of order two, i.e.
\be
\hat{\cal H}_{q=2}[J] = i \sum_{j<k} J_{jk}  \hat{\chi}_j \hat{\chi}_k. 
\ee
where
\be
\{\hat{\chi}_j,\hat{\chi}_k\} = \delta_{jk}.
\ee
Then the square of this Hamiltonian is given by
\bea
\left ( \hat{\cal H}_{q=2}[J] \right )^2&=& - \sum_{j< k, l<m} J_{jk} J_{lm}  \hat{\chi}_j \hat{\chi}_k \hat{\chi}_l \hat{\chi}_m \notag\\
&=& - \hat{I} \sum_{j < k} J_{jk}^2 - 2  \sum_{j< k< l<m} \left ( J_{jk} J_{lm} - J_{jl} J_{km} + J_{jm} J_{kl} \right )  \hat{\chi}_j \hat{\chi}_k \hat{\chi}_l \hat{\chi}_m. 
\eea
where all other terms cancel due to anti-commutativity of  $\hat{\chi}_i$ operators. If we choose $J_{jk}$ coefficients to be random variables such that their product $J_{jk} J_{lm}$ is Gaussian, then 
\be
 \hat{\cal H}_{q=4}[J] = \sum_{j< k< l<m} \left ( J_{jk} J_{lm} - J_{jl} J_{km} + J_{jm} J_{kl} \right )  \hat{\chi}_j \hat{\chi}_k \hat{\chi}_l \hat{\chi}_m \label{eq:SYK} 
 \ee
 is equivalent to the original SYK model \cite{SYK, SYK1, SYK2} for products of up to twelve $\hat{\chi}_k$'s operators, by deviates from it for higher order operators. For example, 
 \bea
 \langle (J_{12} J_{34} - J_{13} J_{24} + J_{14} J_{23}) (J_{12} J_{56} - J_{15} J_{26} + J_{16} J_{25}) (J_{34} J_{56} - J_{35} J_{46} + J_{36} J_{45})  \rangle \notag  \\
 = J_{12}^2 J_{34}^2 J_{56}^2 \neq 0
 \eea
 but the corresponding term in the SYK model must vanish.  
 
 The disorder-averaged partition function of the model,
 \bea
 z[0,\beta', J] &=&\left  \langle tr\left [ \exp\left (\beta' \hat{\cal H}_{q=4}[J] \right ) \right  ] \right \rangle \\\notag 
 & =&  \left  \langle tr\left [ \exp\left (\beta' \sum_{j< k< l<m} \left ( J_{jk} J_{lm} - J_{jl} J_{km} + J_{jm} J_{kl} \right )  \hat{\chi}_j \hat{\chi}_k \hat{\chi}_l \hat{\chi}_m \right ) \right  ]  \right \rangle,
 \eea
 can be obtained by solving the differential equation \eqref{eq:the-equation} for extended partition function,
 \be
 \left (\frac{\partial^2}{ \partial \beta^2} + \sum_{j < k} J_{jk} ^2  + 2 \frac{\partial}{\partial \beta'}  \right ) z[\beta, \beta', J] = 0
 \ee
 with initial conditions
 \bea
 z[\beta, 0, J] &=& \left \langle  tr\left [ \exp\left (\beta \hat{\cal H}_{q=2}[J] \right ) \right  ]  \right \rangle \\\notag
  &=& \left \langle tr\left [ \exp\left (i \beta \sum_{j<k} J_{jk}  \hat{\chi}_j \hat{\chi}_k \right ) \right  ] \right \rangle.
 \eea

\section{Duality pseudo-forest}\label{sec:forest}

An important observation is that the set of the combined operators  $\hat\Gamma^{ij}$'s (defined in \eqref{eq:new_operators}) satisfies the very same conditions that are satisfied by the original operators $\hat\Gamma^{i}$'s. Indeed, the combined operators are (a) Hermitian (due to \eqref{eq:Hermitian})
\be
\left ( \hat\Gamma^{ij} \right )^\dagger = \left ( \hat\Gamma^{i} \hat\Gamma^{j} \right )^\dagger = \hat\Gamma^{j\dagger} \hat\Gamma^{i\dagger} = \hat\Gamma^{j} \hat\Gamma^{i} =  \hat\Gamma^{i} \hat\Gamma^{j} = \hat\Gamma^{ij},
\ee
(b) square to identity (due to \eqref{eq:square} )
\be
\left ( \hat\Gamma^{ij}\right)^2 = \left ( \hat\Gamma^{i} \hat\Gamma^{j}\right)^2 = \left ( \hat\Gamma^{i} \right )^2 \left ( \hat\Gamma^{j}\right)^2 =\hat{I}^2 = \hat{I}, 
\ee
and any pair of them either commutes or anti-commutes or in other words they satisfy the (c) commutativity condition with adjacency matrix defined as
\bea
{ {\cal A}^{ij,kl} }=\left ( {\cal A}^{ik} +   {\cal A}^{il } +  {\cal A}^{jk} +   {\cal A}^{jl} \right ) \mod 2.
\eea
Moreover, if we take an ordered product of the combined operators 
\be
\hat\Gamma^{i_1j_1} \hat\Gamma^{i_2j_2} ... \hat\Gamma^{i_n j_n} = \hat\Gamma^{i_1}\hat\Gamma^{j_1} \hat\Gamma^{i_2}\hat\Gamma^{j_2} ... \hat\Gamma^{i_n}\hat\Gamma^{j_n} \;\;\text{where} \; N i_1 + j_1 < N i_2 + j_2 <  ... < N i_n + j_n\notag
\ee
and then order the right hand side using commutation and anti-commutation relations and eliminate all of the even powers of gamma operators, then we end up with an expression as in  \eqref{eq:tracelessness} (with possibly a minus sign) which must be traceless. Therefore the new operators must satisfy the very same (d) tracelessness condition 
\be
tr\left [\hat\Gamma^{i_1j_1} \hat\Gamma^{i_2j_2} ... \hat\Gamma^{i_n j_n} \right ] =0  \;\;\;\;\;\text{where} \;\; N i_1 + j_1 < N i_2 + j_2 <  ... < N i_n + j_n.
\ee

In Sec. \ref{sec:extended} we argued that ${\cal Z}$ can be always obtained from ${\cal Z}' $, but since the combined  operators in ${\cal Z}' $ satisfies all of the desired properties we can keep going and define ${\cal Z}''$,${\cal Z}'''$, etc. In general starting from a partition function ${\cal Z}_{(1)}$ we would generate a chain of transformations defined by
\be
{\cal  Z}_{(n)}[\beta_{(n)}, J]  =\cosh\left(\beta_{(n)}  \sqrt{ \sum_i J_i^2  +  2 \frac{\partial}{\partial ( \beta_{(n+1)})} } \right ) {\cal Z}_{(n+1)}[\beta_{(n+1)}, J_i J_j]_{\beta_{(n)}=0}
\ee
or in terms of the extended partition function
\be
 \left (\left (\frac{\partial}{ \partial\beta_{(n)}} \right )^2 - \sum_i J_i^2  - 2 \frac{\partial}{\partial \beta_{(n+1)}}  \right ) z[\beta_{(n)}, \beta_{(n+1)} , J] = 0 
\ee
where
\bea
{\cal  Z}_{(n)}[\beta_{(n)}, J_i]  &=&  z[\beta_{(n)}, 0 , J_i]  \\
{\cal Z}_{(n+1)}[\beta_{(n+1)}, J_i J_j] &=&  z[0, \beta_{(n+1)} , J_i].
\eea
This chain either terminates at a constant partition function,
\be
 {\cal  Z}_{(T)}[\beta_{(T)}, J] = \text{const}\label{eq:terminates} 
 \ee
 (i.e. when the set of $\hat\Gamma^{ij} $ operators is empty for some $T$), or enters a closed cycle of transformations
\be
 {\cal  Z}_{(n)}[\beta, J] = {\cal  Z}_{(n+L)}[\beta, J] \label{eq:loop}
\ee
for some $L$ (and $n$ sufficiently large).  For example, the chain of transformations terminates at $T=2$ in the example of Fig. \ref{fig:complete} and enters a closed cycle of self-duality transformations with $L=1$ in the example of Fig. \ref{fig:selfdual}. 

\begin{figure}[]
\begin{center}
\includegraphics[width=1\textwidth]{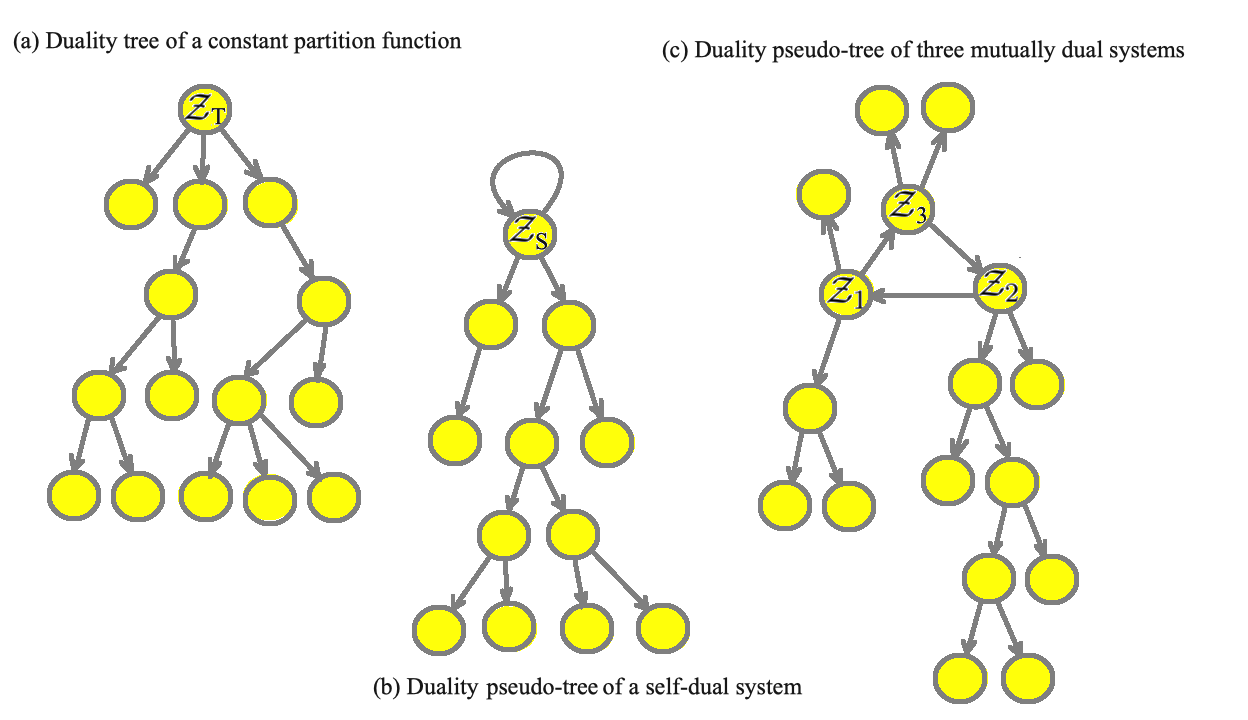}
\caption{Duality pseudo-forest graph with arrows pointing from a parent system to a child system. } \label{fig:forest}
\end{center}
\end{figure}

If we now consider {\it all} quantum Hamiltonians \eqref{eq:Hamiltonian} with gamma operators described by generalized Pauli operators \eqref{eq:Pauli} (for which the tracelessness condition \eqref{eq:tracelessness} is satisfied), and then view the respective quantum systems as vertices of a graph and the duality transformations \eqref{eq:the-equation} as directed edges, then the resulting graph is a directed pseudo-forest, i.e. a graph in which every vertex has at most one incoming edge (see Fig. \ref{fig:forest}). The pseudo-forest consists of a single tree connected to a constant partition function described by Eq. \eqref{eq:terminates} (see Fig. \ref{fig:forest}.a), many pseudo-trees connected to self-dual systems described by Eq. \eqref{eq:loop} with $L=1$ (see Fig. \ref{fig:forest}.b), and all other pseudo-trees connected to closed cycles of transformations between mutually dual systems described by Eq. \eqref{eq:loop} with $L>1$ (see Fig. \ref{fig:forest}.c). For all systems in the tree connected to a constant partition function, the partition functions can be easily obtained by solving a sequence of the second order differential equations with known initial conditions. In this sense all of such systems are exactly solvable. However, for systems in the pseudo-trees connected to a self-dual system (or to cycles of mutually dual systems) the partition functions can be ``easily'' computed only after the partition function of the respective self-dual system (or any of the mutually dual systems) was computed (or at least estimated). In this sense the partition functions of the self-dual systems and the partition functions in the cycles of mutually dual systems are the fundamental structures and the success of solving other systems hinges on our ability to compute (or estimate) these fundamental structures, i.e. partition functions. Nonetheless, to accomplish such an ambitious task one might have to employ iterative numerical methods as was suggested in Sec. \ref{sec:examples} in context of the example of Fig. \ref{fig:selfdual}.

\section{Discussion}\label{sec:discussion}

In this paper we achieved two main results: derived a differential equation \eqref{eq:the-equation}  for the extended partition function (see Sec. \ref{sec:extended}) and then used it to construct a pseudo-forest of duality transformations (see Sec. \ref{sec:forest}). The first result should be viewed as a step towards developing an  operator representation of partition functions (see Sec. \ref{sec:introduction}), while the second result can perhaps be viewed as a  step towards identifying a complete {\it  web} of duality transformations between {\it all} quantum systems. Indeed, the differential operator which annihilates the extended partition function in equation \eqref{eq:the-equation} has a very specific form and as a result (we suspect) only a tiny fraction of {\it all} duality transformation was identified. This might be a good start, but it is also important to study possible applications and generalizations.

For example, consider a Laplace-transformed version of the differential equation \eqref{eq:the-equation},
\be
 \left (\frac{\partial^2}{ \partial \beta^2} - \sum_i \frac{\partial^2}{{\partial x^i}^2}  - 2 \frac{\partial}{\partial \beta'}  \right ) \tilde{z}[\beta, \beta', x^i] = 0 \label{eq:spacetime}
\ee
expressed in terms of the Laplace transform of the extended partition function, i.e.
\be
\tilde{z}[\beta, \beta', \vec{x}] =  \int d^K J \; z[\beta,\beta', J]\; e^{- \sum_i J_i x^i}.
\ee
This equation \eqref{eq:spacetime} can be rewritten in a more symmetric form, 
\be
\left ( \eta^{\mu\nu} \frac{\partial}{ \partial x^\mu}\frac{\partial}{ \partial x^\nu} + 2 \frac{\partial}{\partial \beta'}\right )  \tilde{z}[x^0,  \beta', \vec{x}] = 0\label{eq:spacetime2}
\ee
where $x^0 \equiv \beta$, $\eta=\text{diag}(-1, +1, +1,...)$ and summation over repeated indices is implied. Then one can think of $x^0$ as an emergent time, $\eta_{\mu\nu}$ as an  emergent space-time metric tensor, and $\beta'$ as an emergent extra dimension. For example, if the parent partition function is constant, i.e. $ {z}[0, \beta', \vec{x}]=1$, then  $\frac{\partial}{\partial \beta'}$ drops out and we are left with a massless Klein-Gordon equation
\be
 \eta^{\mu\nu} \frac{\partial}{ \partial x^\mu}\frac{\partial}{ \partial x^\nu}   \tilde{z}[x^0, \vec{x}] = 0\label{eq:spacetime3}
\ee
with a delta-function initial condition
\bea
\tilde{z}[0, \vec{x}] &=& \delta^{(K)}(\vec{x}) \\
\frac{\partial}{\partial \beta} \tilde{z}[0, \vec{x}] &=& 0.
\eea
However, if we are to look for more general operator representations of the partition function (and thus more general duality transformations), then one might try to explore possible generalizations of equation \eqref{eq:spacetime2}. For example, if in equation \eqref{eq:spacetime2} we replace $\eta^{\mu\nu}$ with a non-flat metric tensor $g^{\mu\nu}$ (and partial derivatives with covariant derivatives)  would its solutions $\tilde{z}[x^0, \beta', \vec{x}]$ still describe some other extended partition functions and thus some other duality transformations between quantum systems, i.e. ${z}[0, \beta', \vec{x}]  \rightarrow {z}[x^0, 0, \vec{x}]$? And if so does the emergent space-time has anything to do with the space-time we live in? We leave these as well as other related questions for future work.\\

{\it Acknowledgments.} The author wishes to acknowledge the hospitality of the Tufts Institute of Cosmology where much of the work in completing the paper was carried out. The work was supported in part by the Foundational Questions Institute (FQXi).


\begin{thebibliography}{10}

\bibitem{Zee}
A.~ Zee (2010), ``Quantum Field Theory in a Nutshell,'' Princeton University Press.

\bibitem{Shifman}
 M.~Shifman (2012), ``Advanced Topics in Quantum Field Theory,'' Cambridge University Press.

\bibitem{Kamenev }
A.~Kamenev  (2011) , ``Field Theory of Non-Equilibrium Systems'', Cambridge University Press.

\bibitem{Maldacena}
  J.~M.~Maldacena,
  ``The Large N limit of superconformal field theories and supergravity,''
  Int.\ J.\ Theor.\ Phys.\  {\bf 38}, 1113 (1999)

\bibitem{Witten} 
  E.~Witten,
  ``Anti-de Sitter space and holography,''
  Adv.\ Theor.\ Math.\ Phys.\  {\bf 2}, 253 (1998)


\bibitem{Vanchurin}
  V.~Vanchurin,  ``A quantum-classical duality and emergent space-time,''
  arXiv:1903.06083 [hep-th].
  
\bibitem{Polchinski}
Polchinski, J. (1984), ``Renormalization and Effective Lagrangians'', Nucl. Phys. B, 231 (2): 269,

\bibitem{Wetterich}
Wetterich, C. (1993), ``Exact evolution equation for the effective potential'', Phys. Lett. B, 301 (1): 90,

\bibitem{Morris}
Morris, T. R. (1994), ``The Exact renormalization group and approximate solutions'', Int. J. Mod. Phys. A, A (14): 2411-2449,

\bibitem{SYK}
S. Sachdev and J. Ye, ``Gapless spin fluid ground state in a random, quantum Heisenberg magnet,'' Phys. Rev. Lett. 70 (1993) 3339

\bibitem{SYK1} 
A.~Kitaev, KITP talk, 2015, http://online.kitp.ucsb.edu/online/joint98/kitaev/

\bibitem{SYK2}
  J.~Maldacena and D.~Stanford, ``Remarks on the Sachdev-Ye-Kitaev model,''
  Phys.\ Rev.\ D {\bf 94}, no. 10, 106002 (2016)
  
\bibitem{KW}
H. A. Kramers and G. H. Wannier (1941), ``Statistics of the two-dimensional ferromagnet,'' Physical Review. 60: 252-262.


\end{thebibliography}
\end{document}